# Title: Quantum CNOT Gate for Spins in Silicon

**Authors:** D. M. Zajac[1], A. J. Sigillito[1], M. Russ[2], F. Borjans[1], J. M. Taylor[3], G. Burkard[2], and J. R. Petta[1*]

**Affiliations:**
[1]Department of Physics, Princeton University, Princeton, NJ 08544, USA
[2]Department of Physics, University of Konstanz, D-78457 Konstanz, Germany
[3]Joint Quantum Institute and Joint Center for Quantum Information and Computer Science, NIST and University of Maryland, College Park, MD 20742, USA

*Correspondence to: petta@princeton.edu

**Abstract**: Single qubit rotations and two-qubit CNOT operations are crucial ingredients for universal quantum computing. While high fidelity single qubit operations have been achieved using the electron spin degree of freedom, realizing a robust CNOT gate has been a major challenge due to rapid nuclear spin dephasing and charge noise. We demonstrate an efficient resonantly-driven CNOT gate for electron spins in silicon. Our platform achieves single-qubit rotations with fidelities >99%, as verified by randomized benchmarking. Gate control of the exchange coupling allows a quantum CNOT gate to be implemented with resonant driving in ~200 ns. We use the CNOT gate to generate a Bell state with 75% fidelity, limited by quantum state readout. Our quantum dot device architecture opens the door to multi-qubit algorithms in silicon.

**Main Text:** Gate defined semiconductor quantum dots are a powerful platform for isolating and coherently controlling single electron spins (*1, 2*). Silicon quantum dots can leverage state-of-the-art industrial nanofabrication capabilities for scalability, and support some of the longest quantum coherence times measured in the solid-state (*3-5*). By engineering local magnetic field gradients, electron spins can be electrically controlled (*6, 7*) with single qubit gate fidelities exceeding 99% (*8*). Despite this progress, demonstrations of two-qubit gates with quantum dot spins are scarce due to technological and materials challenges (*9, 10*). While exchange control of spins was demonstrated as early as 2005, high fidelity exchange gates have been difficult to achieve due to nuclear spin dephasing and charge noise (*10, 11*). A demonstration of an efficient CNOT gate for spins in silicon will open a path for multi-qubit algorithms in a scalable semiconductor system.

Here we demonstrate a ~200 ns CNOT gate in a silicon semiconductor double quantum dot (DQD), nearly an order of magnitude faster than the previously demonstrated composite CNOT gate (*9*). The gate is implemented by turning on an exchange interaction, which results in a state-selective electron spin resonance (ESR) transition that is used to implement a CNOT gate with a single microwave (MW) pulse. Local magnetic field gradients allow for all-electrical control of the spin states with single qubit gate fidelities exceeding 99%, enabled by the largely nuclear-spin-free environment of the silicon host lattice. In contrast with previous DQD implementations of the exchange gate, our CNOT gate is implemented at a symmetric operating point, where the exchange coupling $J$ is first-order insensitive to charge noise (*12, 13*). By combining the CNOT with single qubit gates we create a Bell state with a fidelity $F = 75\%$, limited primarily by the qubit readout visibility (*14*). Our demonstration of a universal set of fast quantum gates for spins in silicon paves the way for the first multi-qubit algorithms with semiconductor spin qubits (*15*).

The spin of a single electron is used to encode a qubit (*16*). A gate-defined DQD (Fig. 1A) is used to isolate two electrons in a silicon quantum well, forming a two-qubit device (Fig. 1B). Gate L (R) is used to control the energy of the electron trapped in the left (right) quantum dot, and gate M provides control of $J$. The charge occupancy of the DQD is detected by monitoring the current $I_s$ or conductance $g_s$ through a nearby quantum dot charge sensor (*15*). A Co micromagnet (*17*) generates a magnetic field gradient that results in distinct ESR transition frequencies for the left and right qubits; it also enables high fidelity electrically driven single qubit rotations (*6-8, 18*).

The DQD gate voltages $V_L$ and $V_R$ are rapidly tuned to traverse the charge stability diagram. Starting from the charge state ($N_L = 0$, $N_R = 0$), where $N_L$ ($N_R$) refers to the number of electrons in the left (right) dot, we navigate from points A to C (see Fig. 1C) to initialize the device in the $|\downarrow_L\downarrow_R\rangle$ state. The gates are then pulsed to point D, in the (1, 1) charge state, where single qubit control is achieved by applying MW pulses to gate S. Exchange can be rapidly turned on and off by adjusting the voltage $V_M$. Qubit readout is achieved by moving from points E to G, which sequentially measure and empty the left and right dot spins. Spin dependent tunneling and charge state readout are used to extract the spin-up probability $P_\uparrow^L$ ($P_\uparrow^R$) of the left (right) qubits following pioneering work by Elzerman *et al.* (*14, 19*). Energy level diagrams corresponding to each point in the pulse sequence are shown in Fig. 1D.

We demonstrate high fidelity single qubit control and characterize gate errors using Clifford randomized benchmarking (*20-22*). Figures 2A and B show Rabi oscillations from each qubit, as extracted by measuring the spin-up probabilities as a function of the MW drive frequency and pulse length (with $J = 0$). The ESR frequencies differ by approximately 200 MHz due to the micromagnet field gradient. Driving the right qubit on resonance, we obtain Rabi oscillations that persist for at least 10 µs (Fig. 2C). The average probability differs from 0.5 due to the asymmetric readout fidelities of the spin-up and spin-down states (*17*). Through Ramsey and Hahn echo measurements we find $T_2^* = 1.2$ µs ($T_2^{echo} = 22$ µs) for the left qubit and $T_2^* = 1.4$ µs ($T_2^{echo} = 80$ µs) for the right qubit. Averaging over sequences containing the 24 Clifford gates, randomized benchmarking yields single qubit fidelities $F_L = 99.3 \pm 0.2\%$ and $F_R = 99.7 \pm 0.1\%$.

Proposals for two-qubit interactions with spins in semiconductors are generally based on control of the exchange coupling (*16*). In order to implement a high-fidelity CNOT gate we must first measure $J$ as a function of $V_M$ (*17*). Physically, in the presence of a strong magnetic field gradient $\delta B \gg J$, the exchange interaction lowers the energy of the antiparallel spin states relative to the $|\uparrow\uparrow\rangle$ and $|\downarrow\downarrow\rangle$ spin states (Fig. 3A). As a result, the ESR frequency of the left qubit will be dependent on the state of the right qubit (and vice versa). We can therefore determine $J$ by measuring the left qubit ESR spectra for different right qubit states (Fig. 3B). Specifically, the system is prepared in $|\downarrow\downarrow\rangle$ and then a rotation of duration $\tau_R$ is applied to the right qubit. Next we apply a low power probe tone for a time $\tau_L \gg T_2$ at a frequency $f_p$ that will leave the qubit in a mixed state if $f_p$ is resonant with the qubit frequency. For the simple case where $\tau_R$ is such that the right qubit ends in the spin-down (spin-up) state, the left qubit will have a transition frequency $f^L_{|\psi_R\rangle=|\downarrow\rangle}$ ($f^L_{|\psi_R\rangle=|\uparrow\rangle}$) as illustrated in the green (blue) box in Fig. 3B. By plotting $P^L_\uparrow$ as a function of $\tau_R$ and $f_p$ (Fig. 3C) we see that the left qubit resonance frequency is correlated with the state of the right qubit (Fig. 3D). The exchange frequency $J/h = f^L_{|\psi_R\rangle=|\uparrow\rangle} - f^L_{|\psi_R\rangle=|\downarrow\rangle}$, where $h$ is Planck's constant, is directly extracted from the data sets in Fig. 3E and plotted as a function of $V_M$ in Fig. 3F (*17*). A 20 mV change in $V_M$ is sufficient to turn on a 10 MHz exchange splitting, which exceeds typical single qubit Rabi frequencies ($f_{Rabi} = 4.8$ MHz in Fig. 2C).

Fast gate voltage control of $J$ can be used to implement a resonant CNOT gate (Fig. 4). The general quantum circuit, and its experimental implementation, are shown in Figs. 4A and B. When $V_M$ is low, $J$ is approximately zero ($J \sim 300$ kHz for $V_M = 390$ mV, see Fig. 3F) corresponding to the level diagram on the left in Fig. 4C (*17*). With $J \sim 0$ high fidelity single qubit gates can be implemented, since the resonance frequency of each qubit is independent of the state of the other qubit. When $V_M$ is pulsed high the antiparallel spin states are lowered in energy by $J/2$ relative to the parallel spin states (right panel, Fig. 4C).

To calibrate the CNOT gate, we use a long dc exchange pulse of $\tau_{dc}=1$ µs and vary the length $\tau_P$ of the MW pulse to drive transitions between $|\downarrow\uparrow\rangle$ and $|\uparrow\uparrow\rangle$. Here $|tc\rangle$ describes a product state of the target (t) and control (c) qubits. The resulting conditional oscillations are shown in Fig. 4D for the input states $|\downarrow\uparrow\rangle$ and $|\downarrow\downarrow\rangle$. A conditional π-rotation is realized on the target qubit for $t_{CNOT} = \tau_P = 130$ ns. Due to the magnetic field gradient, changes in $V_M$ shift the orbital positions of the electrons and result in small changes in the ESR resonance frequencies. By setting $\tau_{dc} = 2\pi/J = 204$ ns we eliminate conditional phases due to exchange and the remaining single qubit phases are accounted for in the phase of the consecutive MW drives, resulting in a pure CNOT gate (*17*). In contrast to our single-step CNOT gate, implementation of a conventional CNOT gate following the Loss-DiVincenzo proposal would require mastery of two operations and three single qubit gates, with much more experimental overhead (*16*).

In general, the CNOT gate must be able to operate on an arbitrary input state, and specifically on product states of the form $|\psi_{in}\rangle = (\alpha_L|\downarrow\rangle_L + \beta_L|\uparrow\rangle_L) \otimes (\alpha_R|\downarrow\rangle_R + \beta_R|\uparrow\rangle_R)$, where $R$ and $L$ denote the right (control) and left (target) qubits. Here $|\alpha_{L,R}|^2 + |\beta_{L,R}|^2 = 1$. We now demonstrate that a CNOT gate is produced by first initializing the system in $|\downarrow\downarrow\rangle$. The control qubit is then rotated by an angle $\theta_R$ to create the input state $|\psi_{in}\rangle = |\downarrow\rangle_L \otimes (\cos(\theta_R/2)|\downarrow\rangle_R - i\sin(\theta_R/2)|\uparrow\rangle_R)$. Figure 4E shows $P_\uparrow^L$ and $P_\uparrow^R$ measured after the CNOT gate acts on different input states with angle $\theta_R$. These data show that the target qubit follows the state of the control qubit, as needed for a universal CNOT gate.

We next use the CNOT gate (*17*) to create the Bell state $|\psi_{target}\rangle = \frac{1}{\sqrt{2}}(|\downarrow\downarrow\rangle - i|\uparrow\uparrow\rangle)$. The Bell state fidelity is extracted by performing two qubit state tomography (*23, 24*). By appending single qubit rotations after the CNOT we measure the expectation value for all two qubit Pauli operators (for example, by applying a $\pi/2_x$ rotation to the left qubit and $\pi/2_y$ rotation to the right we measure the YX two qubit operator). Since the set of Pauli operators form a basis of the Hermitian operators on the two qubit Hilbert space we can reconstruct the full two qubit density matrix from these measurements. The fidelity of the prepared Bell state is $F = \sqrt{\langle\psi_{target}|\rho|\psi_{target}\rangle} = 75\%$ (*25*). The readout visibilities of both qubits (*17*), and spin relaxation during the sequential qubit readout, account for 20% of the fidelity loss. We anticipate that the readout fidelity could be improved using cold amplifiers or dispersive measurement techniques (*26, 27*).

Realizing robust two-qubit gates has been a bottleneck in the development of spin-based quantum computers (*10*). The silicon DQD system presented here allows for rapid control of the nearest neighbor exchange coupling and enables an efficient resonantly driven CNOT gate in ~200 ns, nearly an order of magnitude faster than previous results and on par with single qubit gate operation times in Si (*9*). A Bell state is generated using the CNOT gate, with a state fidelity $F = 75\%$ that is largely limited by the readout visibility. By leveraging technological improvements in semiconductor DQD readout to increase the measurement visibility (*27*), demonstrations of simple quantum algorithms with a silicon spin processor may be within reach.

**Acknowledgments:** We thank T. Hazard, J. Stehlik, and K. Wang for technical assistance. Research was sponsored by Army Research Office grant W911NF-15-1-0149, the Gordon and Betty Moore Foundation's EPiQS Initiative through grant GBMF4535, and NSF grant DMR-1409556. Devices were fabricated in the Princeton University Quantum Device Nanofabrication Laboratory.

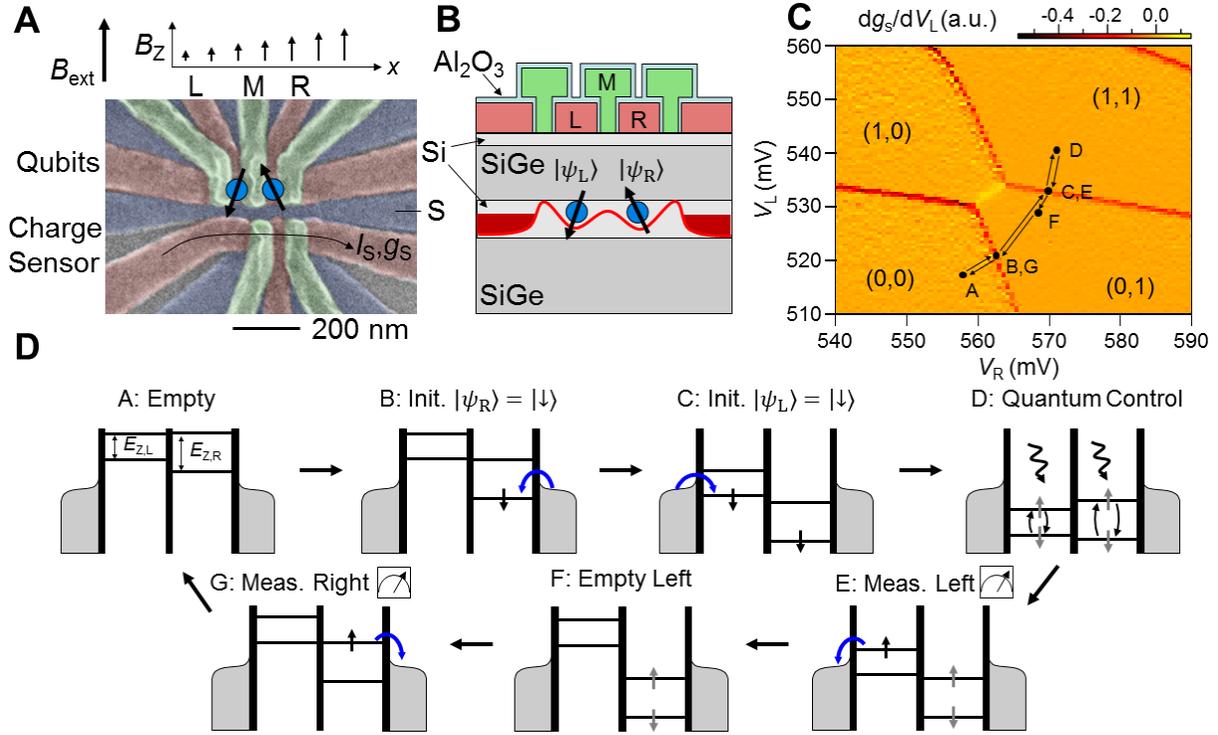

**Fig. 1. Two-qubit device.** (**A**) False-color scanning electron microscope image of the DQD before deposition of the Co micromagnet. Two spin qubits are defined by the DQD and a neighboring quantum dot is used as a charge sensor. The Co micromagnet (not shown) creates a slanting Zeeman field that is used for quantum control. (**B**) Schematic cross-section of the DQD device. Two electrons are trapped in the confinement potential created by gates L, M, and R. (**C**) DQD charge stability diagram. Points A-G are used in the two-qubit control sequence. (**D**) DQD energy level configuration at each point in the pulse sequence. Points A-C are used to initialize the system in $|\downarrow\downarrow\rangle$. Single qubit and two qubit gates are implemented at point D. Sequential single-shot spin state readout is achieved by navigating from points E-G.

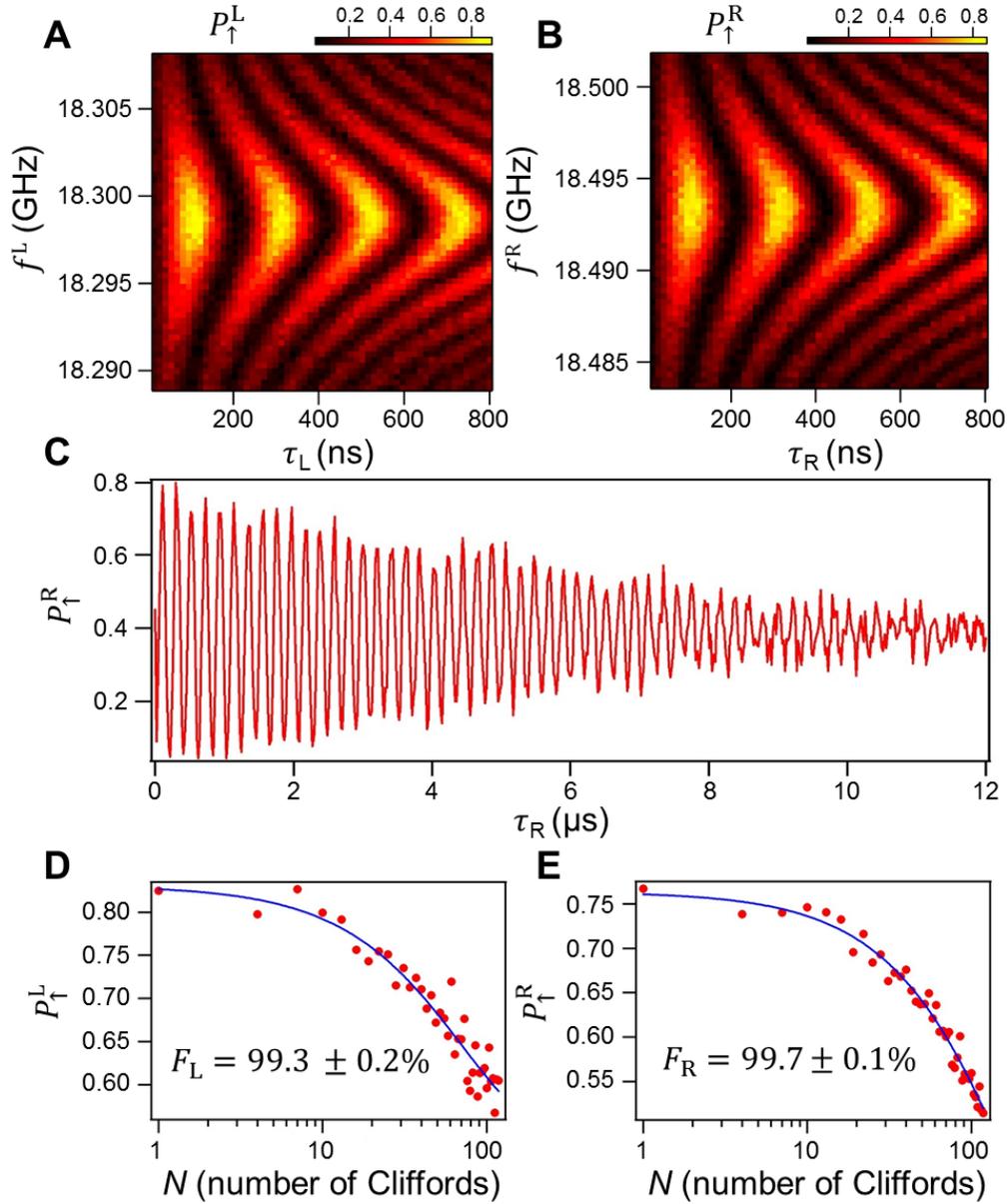

**Fig. 2. High fidelity single qubit gates.** (**A**) Left qubit spin-up probability $P_\uparrow^L$ plotted as a function of the MW drive frequency $f^L$ and drive time $\tau_L$, showing coherent Rabi oscillations. (**B**) Right qubit Rabi oscillations. (**C**) $P_\uparrow^R$ as a function of $\tau_R$ shows high visibility Rabi oscillations that persist to 10 μs. Clifford randomized benchmarking of the left (**D**) and right (**E**) qubits yields gate fidelities in excess of 99%.

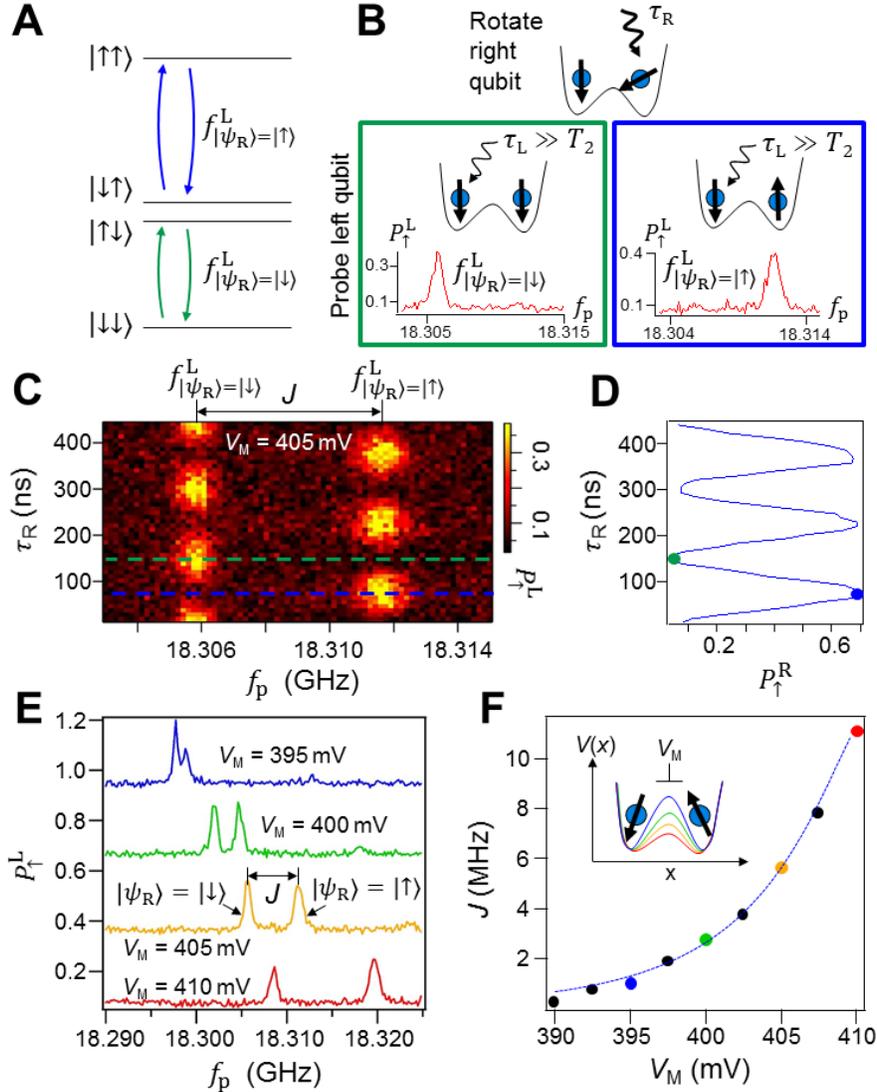

**Fig. 3. Exchange spectroscopy.** (**A**) Schematic energy level diagram with a large $J$. (**B**) We spectroscopically measure $J$ by first applying a rotation to the right qubit and then applying a low power probe tone to the left qubit. The left qubit will have a resonance frequency $f^L_{|\psi_R\rangle=|\downarrow\rangle}$ when the right qubit is in the spin-down state (green box) and $f^L_{|\psi_R\rangle=|\uparrow\rangle}$ when the right qubit is in the spin-up state (blue box). (**C**) $P^L_\uparrow$ as a function of $\tau_R$ and the probe frequency $f_p$. The two resonance frequencies of the left qubit are split by $J$. The response of the left qubit to the probe tone oscillates between these two frequencies as the right qubit oscillates between spin-up and spin-down. (**D**) $P^R_\uparrow$ as a function of $\tau_R$, displaying Rabi oscillations. (**E**) Spectra showing the left dot resonance frequencies for four different values of $V_M$. Curves offset by 0.3 for clarity. (**F**) $J$ as a function of $V_M$ (dots) and theory (line).

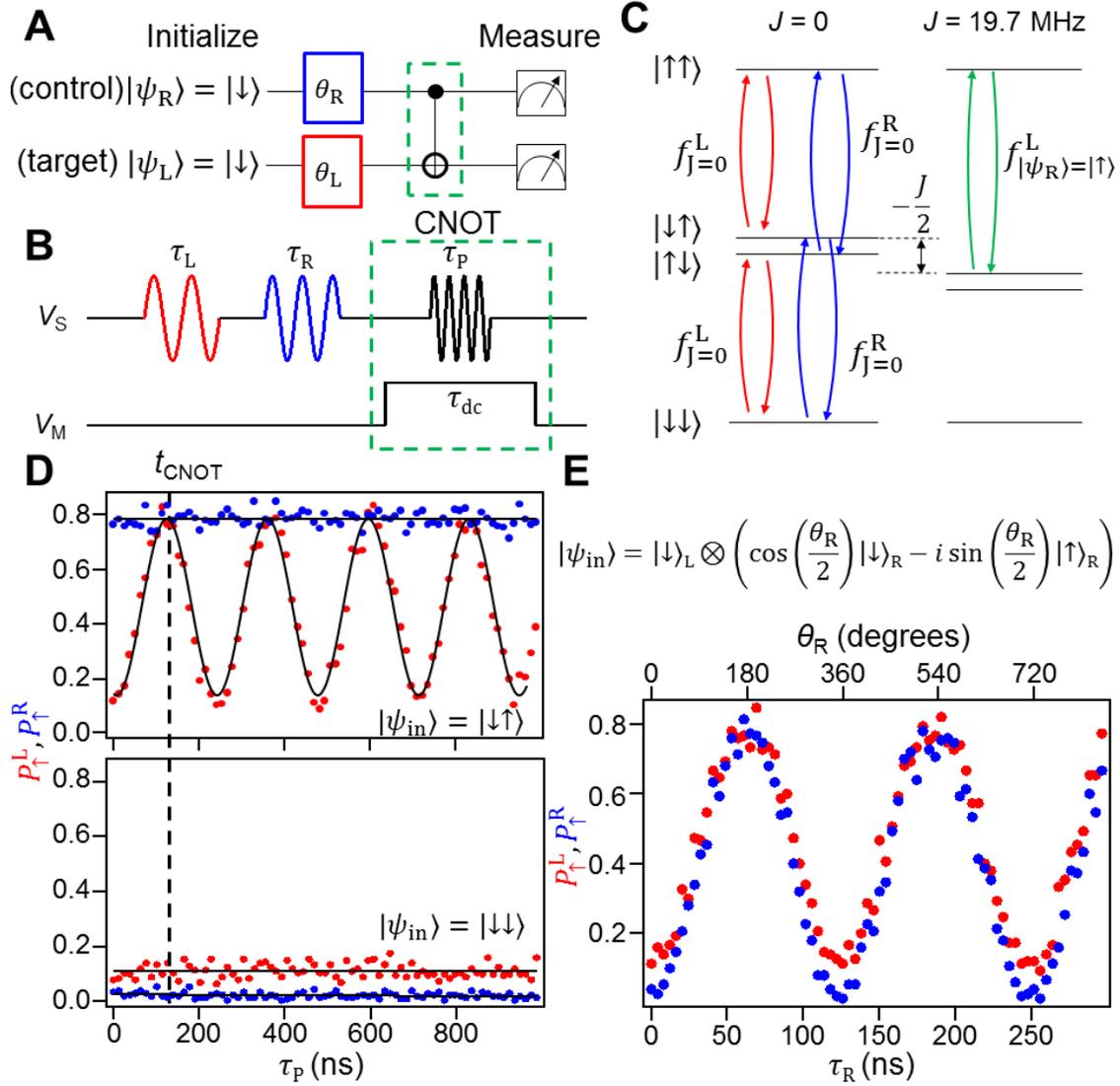

**Fig. 4. CNOT gate.** (**A**) Quantum circuit for the CNOT gate. (**B**) Experimental implementation of the quantum circuit. (**C**) Schematic energy level diagrams for $J=0$ and $J\neq 0$. When $J\neq 0$ a conditional rotation can be applied to the left qubit by driving at $f^L_{|\psi_R\rangle=|\uparrow\rangle}$. (**D**) $P^L_\uparrow$ (red) and $P^R_\uparrow$ (blue) as a function of $\tau_P$ for input states $|\downarrow\uparrow\rangle$ and $|\downarrow\downarrow\rangle$. The vertical dashed line at $\tau_P = 130$ ns $= t_{\text{CNOT}}$ corresponds to a CNOT gate. (**E**) CNOT gate with superposition input states. With $\tau_P = 130$ ns and $\tau_L = 0$, the response of the CNOT gate is plotted as a function of $\tau_R$ showing that the target qubit follows the state of the control qubit.

# Supplementary Materials for

## Quantum CNOT Gate for Spins in Silicon

**Materials and Methods**

The double quantum dot (DQD) device is fabricated from three layers of overlapping aluminum gates on top of a Si/SiGe quantum well heterostructure in natural Si (Figs. 1A and B). A Co micromagnet is deposited on top of the device to provide a slanting Zeeman field for spin manipulation (*1*). Two Agilent 33250 signal generators are used to control $V_L$ and $V_R$ throughout our experiments. Two Agilent 8267D microwave (MW) sources are used to drive gate S for performing single spin manipulation. To prevent unintentional microwave leakage from resonantly driving our system, each source was detuned from the electron spin resonance (ESR) frequency by 30 MHz. Using the Agilent 8267D internal IQ mixers, the baseband microwave frequencies were mixed up to the qubit frequencies using a Tektronix AWG5014 arbitrary waveform generator. A second Tektronix AWG5014C is used for rapid control of the exchange coupling via square pulses applied to gate M.

**Supplementary Text**

1. Micromagnet Design

The micromagnet design is shown in Fig. S1 and is based on recent work by Yoneda *et al.* (*1*). An external field $B_{\text{ext}} = 0.5$ T is applied along the z-axis to magnetize the cobalt film. There are in principle nine different components of the field gradient created by the micromagnet: $dB_i/dj$ where $i, j = x, y, z$. The micromagnet is designed to maximize $dB_z/dx$ and $dB_y/dz$. The $dB_z/dx$ gradient gives rise to the different Zeeman splittings of the two qubits since they are spatially separated along the x-direction. The $dB_y/dz$ gradient is intended to be the main mechanism for driving spin rotations. Using the voltage on gate S we oscillate the position of the electrons primarily in the z-direction, leading to an oscillating $B_y$ magnetic field.

2. DQD Stability Diagram

The ground state charge configuration of the DQD is mapped out by monitoring the conductance of the charge sensor $g_s$ as a function of the left and right plunger gate voltages $V_L$ and $V_R$. The resulting charge stability diagram is shown in Fig. S2 where we plot the derivative of $g_s$ with respect to $V_L$. The DQD is highly tunable and the (0,0) charge state is reached in the lower left corner of the charge stability diagram.

3. Hamiltonian of a DQD in a Slanting Field

We operate our device in a regime where the difference of the Zeeman splitting in the two dots is much larger than the strength of the exchange interaction $\delta E_Z \gg J$ during all parts of the experiment. In this regime the exchange interaction will not drive a conventional SWAP operation due to the Zeeman energy difference of the left and right spins (2). Instead, the resonance frequency of each qubit becomes dependent on the state of the other qubit. We make use of this conditional single qubit frequency to directly drive a CNOT operation.

The DQD can be described in the basis $\{|\uparrow\uparrow\rangle, |\uparrow\downarrow\rangle, |\downarrow\uparrow\rangle, |\downarrow\downarrow\rangle\}$ by the time dependent Hamiltonian

$$H(t) = J(t)\left(\mathbf{S}_L \cdot \mathbf{S}_R - \frac{1}{4}\right) + \mathbf{S}_L \cdot \mathbf{B}_L + \mathbf{S}_R \cdot \mathbf{B}_R. \tag{1}$$

Here, the first term in the Hamiltonian describes the exchange interaction with strength $J(t)$ between the spin of the electron in the left dot $\mathbf{S}_L$ and the spin of the electron in the right dot $\mathbf{S}_R$. The second (third) term describes the influence of the magnetic field (in energy units), $\mathbf{B}_L = \left(0, B_{y,L}^m(t), B_z^{ext} + B_{z,L}^m(t)\right)^T$ and $\mathbf{B}_R = \left(0, B_{y,R}^m(t), B_z^{ext} + B_{z,R}^m(t)\right)^T$, on the spin in the left (right) dot, composed of a homogeneous external magnetic field in the z-direction, $B_z^{ext}$ and an inhomogeneous field due to the micromagnet that contributes to both the y-component of each dot $B_{y,L}^m(B_{y,R}^m)$ and the z-component $B_{z,L}^m(B_{z,R}^m)$. We assume here that the magnetic field gradients are smaller than the single dot charging energies: $|\mathbf{B}_L - \mathbf{B}_R| \ll E_{C,L}, E_{C,R}$, where $E_{C,L}, E_{C,R}$ are the charging energies of both dots. In our experiment $E_{C,L}, E_{C,R} \sim 6$ meV and $|\mathbf{B}_L - \mathbf{B}_R| \sim 1$ µeV. Meunier *et al.* present a general Hamiltonian valid outside of this assumption in ref. (3).

We perform single qubit operations in a regime where $J$ is approximately zero. Single-qubit gates are implemented by applying a sinusoidal drive to the S gate (shown in Fig. 1A of the main text). The oscillating electric field causes the positions of both electrons to oscillate in the fringing field of the micromagnet producing time-dependent out of plane magnetic fields in the DQD Hamiltonian:

$$B_{y,L}^m(t) = B_{y,L}^0 + B_{y,L}^1 \cos(2\pi ft + \varphi) \tag{2}$$

$$B_{y,R}^m(t) = B_{y,R}^0 + B_{y,R}^1 \cos(2\pi ft + \varphi) \tag{3}$$

where $B_{y,L}^0$ and $B_{y,R}^0$ are the static components of the out of plane fields determined by the average positions of the electrons, and $B_{y,L}^1$ and $B_{y,R}^1$ are the magnitudes of the oscillating fields determined by the amplitude of the microwave drive applied to gate S. By matching the microwave drive frequency $f$ with the single spin resonance frequencies we can individually address each spin. The single spin rotation axis can be controlled by adjusting the phase $\varphi$ of the microwave signal, producing an x rotation for $\varphi = -\pi/2$ and a y rotation for $\varphi = 0$.

We measure the qubit coherence times using Ramsey and Hahn echo pulse sequences. From the Ramsey decay we extract $T_2^* = 1.2$ µs for the left qubit and $T_2^* = 1.4$ µs for the right qubit. From the Hahn echo measurements we find $T_2^{echo} = 22$ µs for the left qubit and $T_2^{echo} = 80$ µs for the right qubit. The large difference in $T_2^{echo}$ may be due to different field gradients seen by each qubit. A larger field gradient will lead to a stronger coupling of the qubit to charge noise and lead to faster dephasing. Further experiments are needed to elucidate the discrepancy.

## 4. Single Qubit Randomized Benchmarking

In order to assess the single qubit gate fidelity we perform Clifford randomized benchmarking on both qubits (*4-6*). In Clifford randomized benchmarking $N$ Clifford rotations are applied to rotate the spin qubit. At the end of the sequence a rotation is applied that would bring the qubit back to the spin-up state in the absence of any errors. The final spin state is then measured. The spin-up return probability $P_\uparrow = Ap_c^N + B$ is plotted as a function of the number of Cliffords $N$ for both qubits in Figs. 2D and E. The spin-up probability is then fit by $P_\uparrow = Ap_c^N + B$ where $p_c$ is the sequence fidelity. The Clifford fidelity can then be expressed in terms of the sequence fidelity as $F_c = (1 + p_c)/2$. From these fits we obtain a Clifford fidelity of $F_L = 99.3 \pm 0.2$ % and $F_R = 99.7 \pm 0.1$ % for the left and right qubits, respectively.

## 5. Readout Fidelity from Numerical Simulations

To understand the limitations in our measurement fidelity and readout visibility, we have constructed a model similar to that used by Morello *et al.* (*7*). Our model can be summarized as follows:
1. Generate a series of time-domain current traces based on the measured tunneling rates $\Gamma^{on}$ and $\Gamma^{off}$ of both quantum dots.
2. Incorporate relaxation and loading errors by modifying an appropriate number of current traces.
3. Apply noise to the current traces using the measured noise spectrum.
4. Filter the current traces using the same software filters applied during the experiment and histogram the maximum currents.

### 5.1 Determining Tunneling Rates

To determine the tunneling rates, both qubits were prepared in the spin-up state and sequential single shot spin readout was performed 5000 times using the technique described by Elzerman *et al.*, (*8*). The delay between the start of the readout and the rising edge of the current pulse is governed by the rate at which electrons tunnel off the quantum dot ($\Gamma^{off}$), so by constructing a histogram of the rising edge times, one can fit the data to an exponential with a characteristic time constant $\Gamma^{off}$. Similarly, the falling edge of a given current pulse represents a tunneling-in event, so by constructing a histogram of the width of the current pulses, we can extract the tunneling-in rate $\Gamma^{on}$. These data are plotted in Fig. S3.

### 5.2 Estimating the Readout Fidelity

Using the measured tunneling rates we generate simulated current traces for spin up and spin down electrons without noise. Based on our measured spin relaxation time $T_1 = 22$ ms and the simulated time before the electron hops out $\tau^{off}$ we convert a fraction of the simulated spin up traces to spin down with the following probability to simulate relaxation:

$$P_r = e^{-\tau^{\text{off}}/T_1}.$$

Similarly we convert a fraction of the spin down traces to spin up based on the calculated probability of unintentionally initializing into spin up based on our measured electron temperature $T_e \approx 150$ mK.

Next we add noise to the simulated traces using the measured noise spectrum of the charge sensor. Finally we filter the simulated current traces using the same filter applied in our experiment and histogram the peak-to-peak current $\Delta I_S$ measured during the readout window. The spin up readout fidelity $F_\uparrow$ and spin down readout fidelity $F_\downarrow$ are plotted as a function of the threshold current $I_{\text{th}}$ in Fig. S4. The visibility is also calculated according to $V = 1 - F_\uparrow - F_\downarrow$ and plotted on the same graph. We find that given our experimental setup, our best expected visibility is 85% for the left dot and 78% for the right dot. This analysis is consistent with the maximum visibilities we have achieved in the experiment.

## 5.3 Limitations to the Readout Fidelity

In these experiments, the electron temperature accounts for 10% of the reduction in our visibility, and spin relaxation accounts for the remaining 5-10%. In our experiment we readout the qubits sequentially, reading the left qubit first. Relaxation of the right qubit during the readout of the left qubit is what leads to the asymmetry of the expected visibilities. Relaxation contributions to the readout fidelity can be mitigated by using a faster readout technique such as RF reflectometry or by incorporating a cold amplifier into the readout circuit (*9, 10*). Further improvements can be made by reducing the electron temperature or using cavity-based measurement approaches (*11*).

## 6. Exchange Interaction

The strength of the exchange interaction is experimentally adjusted by precisely controlling the voltage on the middle barrier gate $V_M$. Lowering (raising) the energy barrier between the dots increases (decreases) the transition matrix elements between the wavefunctions in the two dots. Additionally, changing the energy barrier between the dots also changes the position of the electrons in the field gradient thus adding a time-dependent contribution to the z-component of the magnetic fields $B_{z,Q}^m(t) = B_{z,Q}^{m,0} + B_{z,Q}^{m,1}(t)$ with $Q = L, R$. The two parameters $B_{z,L}^{m,1}$ and $B_{z,R}^{m,1}$ are responsible for the single-qubit frequency shifts that occur when exchange is turned on.

Setting $B_{y,L}(t) = B_{y,R}(t) = 0$ and defining the average Zeeman splitting $E_Z = \left(B_z^{\text{ext}} + B_{z,L}^{m,0} + B_z^{\text{ext}} + B_{z,R}^{m,0}\right)/2$ and the splitting between the two qubits $\delta E_Z = \left(B_z^{\text{ext}} + B_{z,R}^{m,0} - B_z^{\text{ext}} - B_{z,L}^{m,0}\right)$ we find for the instantaneous eigenvalues

$$\mathcal{E}(|\uparrow\uparrow\rangle) = E_Z + \left(B_{z,L}^{m,1} + B_{z,R}^{m,1}\right)/2, \tag{4}$$

$$\mathcal{E}(|\widetilde{\uparrow\downarrow}\rangle) = \left(-J - \sqrt{J^2 + \left(J^2 + \left(\delta E_Z - \left(B_{Z,L}^{m,1} - B_{Z,R}^{m,1}\right)\right)\right)^2}\right)/2, \tag{5}$$

$$\mathcal{E}(|\widetilde{\downarrow\uparrow}\rangle) = \left(-J + \sqrt{J^2 + \left(J^2 + \left(\delta E_Z - \left(B_{Z,L}^{m,1} - B_{Z,R}^{m,1}\right)\right)\right)^2}\right)/2, \tag{6}$$

$$\mathcal{E}(|\downarrow\downarrow\rangle) = -E_Z - \left(B_{Z,L}^{m,1} + B_{Z,R}^{m,1}\right). \tag{7}$$

The tilde indicates the hybridization of the $|\uparrow\downarrow\rangle$ and $|\downarrow\uparrow\rangle$ states due to the exchange interaction. These eigenvalues are plotted as a function of $J$ in Fig. S5. As $J$ is increased the antiparallel spin states are lowered in energy with respect to the parallel spin states. We find the corresponding transition frequencies

$$f^L_{|\psi_R\rangle=|\downarrow\rangle} = \left|\mathcal{E}(|\downarrow\downarrow\rangle) - \mathcal{E}(|\widetilde{\uparrow\downarrow}\rangle)\right| = E_Z + \left(-J + B_{Z,L}^{m,1} + B_{Z,R}^{m,1} - \sqrt{J^2 + \left(\delta E_Z - \left(B_{Z,L}^{m,1} - B_{Z,R}^{m,1}\right)\right)^2}\right)/2, \tag{8}$$

$$f^L_{|\psi_R\rangle=|\uparrow\rangle} = \left|\mathcal{E}(|\widetilde{\downarrow\uparrow}\rangle) - \mathcal{E}(|\uparrow\uparrow\rangle)\right| = E_Z + \left(J + B_{Z,L}^{m,1} + B_{Z,R}^{m,1} - \sqrt{J^2 + \left(\delta E_Z - \left(B_{Z,L}^{m,1} - B_{Z,R}^{m,1}\right)\right)^2}\right)/2, \tag{9}$$

$$f^R_{|\psi_L\rangle=|\downarrow\rangle} = \left|\mathcal{E}(|\downarrow\downarrow\rangle) - \mathcal{E}(|\widetilde{\downarrow\uparrow}\rangle)\right| = E_Z + \left(-J + B_{Z,L}^{m,1} + B_{Z,R}^{m,1} + \sqrt{J^2 + \left(\delta E_Z - \left(B_{Z,L}^{m,1} - B_{Z,R}^{m,1}\right)\right)^2}\right)/2, \tag{10}$$

$$f^R_{|\psi_L\rangle=|\uparrow\rangle} = \left|\mathcal{E}(|\widetilde{\uparrow\downarrow}\rangle) - \mathcal{E}(|\uparrow\uparrow\rangle)\right| = E_Z + \left(-J - B_{Z,L}^{m,1} - B_{Z,R}^{m,1} + \sqrt{J^2 + \left(\delta E_Z - \left(B_{Z,L}^{m,1} - B_{Z,R}^{m,1}\right)\right)^2}\right)/2. \tag{11}$$

An important observation is that the shift in either electron's resonance frequency due to the state of the other electron exactly matches the exchange splitting

$$f^{L}_{|\psi_R\rangle=|\uparrow\rangle} - f^{L}_{|\psi_R\rangle=|\downarrow\rangle} = f^{R}_{|\psi_L\rangle=|\uparrow\rangle} - f^{R}_{|\psi_L\rangle=|\downarrow\rangle} = J. \tag{12}$$

## 7. Exchange Spectroscopy

The pulse sequence used for the exchange spectroscopy of Fig. 3 is shown in Fig. S6. Both qubits are first initialized in spin-down. Then a microwave (MW) drive of length $\tau_R$ is applied to the right qubit preparing it in a state $|\psi\rangle_R = (\cos(\theta/2)|\downarrow\rangle_R - i\sin(\theta/2)|\uparrow\rangle_R)$ where $\theta$ is the rotation angle. Next, a low power drive of fixed length $\tau_L \gg T_2$ is applied to the left qubit at frequency $f_p$. This low power probe tone is used to determine the resonance frequency of the left qubit, which will depend on the state of the right qubit due to exchange. When $f_p$ is on resonance with the left qubit it will leave the left qubit in a random state with 50% probability of being spin up. Finally both qubits are read out. The results of these measurements are shown in Figs. 3C and D of the main text. We observe that the left qubit resonance frequency alternates between two values, correlated with the oscillations in the spin state of the right qubit. This demonstrates that the left qubit resonance frequency is determined by the state of the right qubit, and also allows us to measure the exchange interaction as a function of the gate voltage $V_M$.

The exchange splitting can be approximated for $t_c, |\mathbf{B}_L - \mathbf{B}_R| \ll |E_C \pm \epsilon|$ by

$$J = \frac{2t_c^2(E_{C,L}+E_{C,R})}{(E_{C,L}+\epsilon)(E_{C,R}-\epsilon)}, \tag{13}$$

with $E_{C,L}$ and $E_{C,R}$ being the charging energy of the left and right quantum dots, and $t_c$ being the tunneling matrix element between the dots. The tunneling matrix element depends on the middle barrier potential $V_M$ and can be approximated as $t_c = E\tau$ where $\tau$ denotes the tunneling amplitude and $E$ the single-dot confinement energy. Then $J \propto t_c^2 \propto |\tau|^2$. As a simple model, we use the tunneling probability through a rectangular barrier in the limit of weak tunneling

$$|\tau|^2 = \frac{16E(V-E)}{V^2}exp(-2W\sqrt{2m|V-E|/\hbar^2}), \tag{14}$$

where $V$ and $W$ denote the potential barrier height and width and $m$ the electron mass. Using $V \propto -V_M + const.$ we find

$$J(V_M) = c\frac{V_{M0}-V_M}{(V_M-V_{M1})^2}exp(-\sqrt{|V_M-V_{M0}|/V_{on}}), \tag{15}$$

Where $V_{M1}$ is the voltage at which the tunneling barrier height would be zero, $V_{M0}$ is the voltage at which the barrier height would equal the electron energy, $V_{on}$ describes the voltage scale of the sub-exponential increase of $J$ with $V_M$, and $c$ is an overall scale factor. By fitting the experimentally determined values of $J$ to this function we find $V_{on} = 0.559\ V$, $V_{M0} = 412.8\ V$, $V_{M1} = 451.8\ V$ and $c = 16000$ (GHz)$^2$/e. The measured $J(V_M)$ is in good agreement with the predictions of this simple model (Fig. S7).

8. CNOT Gate

A single-shot two-qubit CNOT gate is implemented by applying a square pulse to the M-gate to turn on the exchange interaction for a time $\tau_{dc}$ while simultaneously driving the S-gate at the transition frequency $f^L_{|\psi_R\rangle=|\uparrow\rangle}$ for a time $\tau_P$. The resonant driving results in coherent rotations between the $|\widetilde{\downarrow\uparrow}\rangle$ and $|\uparrow\uparrow\rangle$ states while the other states stay untouched. This results in a CNOT gate with the right qubit as the control and the left qubit as the target.

8.1 DC Exchange Dynamics

We experimentally measure the four transition frequencies when $J$ is nonzero by making use of a Hahn echo sequence (*12*). To measure $f^L_{|\psi_R\rangle=|\downarrow\rangle}$ for instance, we start with the state $|\psi_{in}\rangle = |\downarrow\downarrow\rangle$ and perform a Hahn echo pulse sequence on the left qubit. By placing an exchange pulse in one half of the free evolution period, we induce a single qubit phase on the left dot which is not cancelled out by the echo. Measuring the spin up probability of the left dot as a function of the length of the exchange pulse $\tau_{dc}$, we can track the phase acquired by the left qubit when the right qubit is spin-down (red trace in Fig. S8B). The rate of phase accumulation is determined by $f^L_{|\psi_R\rangle=|\downarrow\rangle}$. Repeating this experiment with the input state $|\psi_{in}\rangle = |\downarrow\uparrow\rangle$ gives a corresponding measurement of $f^L_{|\psi_R\rangle=|\uparrow\rangle}$ (blue trace in Fig. S8B). Similarly we can measure $f^R_{|\psi_L\rangle=|\uparrow\rangle}, f^R_{|\psi_L\rangle=|\downarrow\rangle}$ by applying the echo to the right qubit for both states of the left qubit.

By choosing $\tau_{dc}$ appropriately we can cancel out any conditional phases due to exchange. For a better understanding let us focus only on the phases of the left qubit; preparing both spins in a superposition we start with the initial wavefunction

$$|\psi(0)\rangle = (|\downarrow\rangle_L + |\uparrow\rangle_L) \otimes (|\downarrow\rangle_R + |\uparrow\rangle_R)/2 = (|\downarrow\rangle_L + |\uparrow\rangle_L) \otimes |\downarrow\rangle_R/2 + (|\downarrow\rangle_L + |\uparrow\rangle_L) \otimes |\uparrow\rangle_R/2. \quad (16)$$

Consider the evolution of the wavefunction after the exchange is adiabatically turned on and held at a value $J$ for a time $\tau_{dc}$. Looking at only the phase acquisition on the left dot, the wavefunction after a time $\tau_{dc}$ evolves into

$$|\psi(t)\rangle = \left(|\downarrow\rangle_L + e^{-if^L_{|\psi_R\rangle=|\downarrow\rangle}\tau_{dc}}|\uparrow\rangle_L\right) \otimes |\downarrow\rangle_R/2 + \left(|\downarrow\rangle_L + e^{-if^L_{|\psi_R\rangle=|\downarrow\rangle}\tau_{dc}}e^{-iJ\tau_{dc}}|\uparrow\rangle_L\right) \otimes |\uparrow\rangle_R/2, \quad (17)$$

due to the property $f^L_{|\psi_R\rangle=|\uparrow\rangle} = f^L_{|\psi_R\rangle=|\downarrow\rangle} + J$ from Eq. (12). Thus for $\tau_{dc} = 2\pi/J$ the two-qubit C-phase operation is effectively canceled out so that the left dot will have the same phase regardless of the state of the right dot. The remaining single qubit phase is determined solely by $f^L_{|\psi_R\rangle=|\downarrow\rangle}$. The same argument holds for the phases of the right qubit.

## 8.2 AC Gate

Due to the choice of $\tau_{dc} = 2\pi/J$ the conditional phases accumulated by the dc exchange pulse are completely cancelled out. An additional ac gate is applied to resonantly drive a spin transition between $|\widetilde{\downarrow\uparrow}\rangle$ and $|\uparrow\uparrow\rangle$. This results in a CNOT gate with the following truth table:

$$
\begin{aligned}
|\uparrow\uparrow\rangle &\to |\downarrow\uparrow\rangle, \\
|\uparrow\downarrow\rangle &\to |\uparrow\downarrow\rangle, \\
|\downarrow\uparrow\rangle &\to |\uparrow\uparrow\rangle, \\
|\downarrow\downarrow\rangle &\to |\downarrow\downarrow\rangle.
\end{aligned}
\tag{18}
$$

To support our experiment we numerically simulate the unitary time evolution of our two qubit system. We integrate the time-dependent Schrodinger equation $i\hbar\dot{\Psi}(t) = H(t)\Psi(t)$ in the rotating frame of the two microwave sources $\widetilde{H}(t) = U^\dagger H U + iU^\dagger \dot{U}$ with $U = exp[i\omega t(S_{z,L} + S_{z,R})/\hbar]$ to cancel out the fast oscillations from the time-independent magnetic field gradient terms. In Fig. S9 we numerically simulate the pulse sequence for the frequency selective microwave gate detailed in Fig. 4 of the main text. The spin-up probabilities for the two input states $|\psi_{in}\rangle = |\uparrow\uparrow\rangle$ and $|\psi_{in}\rangle = |\downarrow\uparrow\rangle$ show anti-correlated oscillations. For $\tau_p = 130$ ns a CNOT is realized, in good agreement with the experimental data.

## 9. State Tomography of a Bell State

We now use the CNOT gate to prepare and readout a Bell state. By applying a $\pi/2$ pulse on the right qubit (control) we prepare the input state $|\psi_{in}\rangle = (|\downarrow\downarrow\rangle - i|\downarrow\uparrow\rangle)/\sqrt{2}$. After applying the CNOT gate, the resulting wave function is the Bell state $|\psi_{out}\rangle = (|\downarrow\downarrow\rangle - i|\uparrow\uparrow\rangle)/\sqrt{2}$. We then use single qubit gates to perform full two qubit state tomography on the output state. The measured density matrix is shown in Fig. S10B, and the measured fidelity is $F = \sqrt{\langle\psi_{target}|\rho|\psi_{target}\rangle} = 75$ %. Due to the finite readout visibility of the left $V_L \approx 0.76$ and right qubits $V_R \approx 0.70$ we expect to see significant false counts of the $|\downarrow\uparrow\rangle$ and $|\uparrow\downarrow\rangle$ states reducing the fidelity. Beyond misidentifying the single spin states, the limited visibility also affects the qubit correlation measurements. We expect the correlation visibility to be proportional to the product of the two single spin visibilities. The expected density matrix for the ideal Bell state is shown in Fig. S10A. The fidelity of this ideal density matrix is $F = \sqrt{\langle\psi_{target}|\rho_{ideal}|\psi_{target}\rangle} = 80.5$ % with the reduction in fidelity coming solely from limited readout visibility. This indicates that most of the fidelity is lost due to readout visibility rather than two qubit gate error.

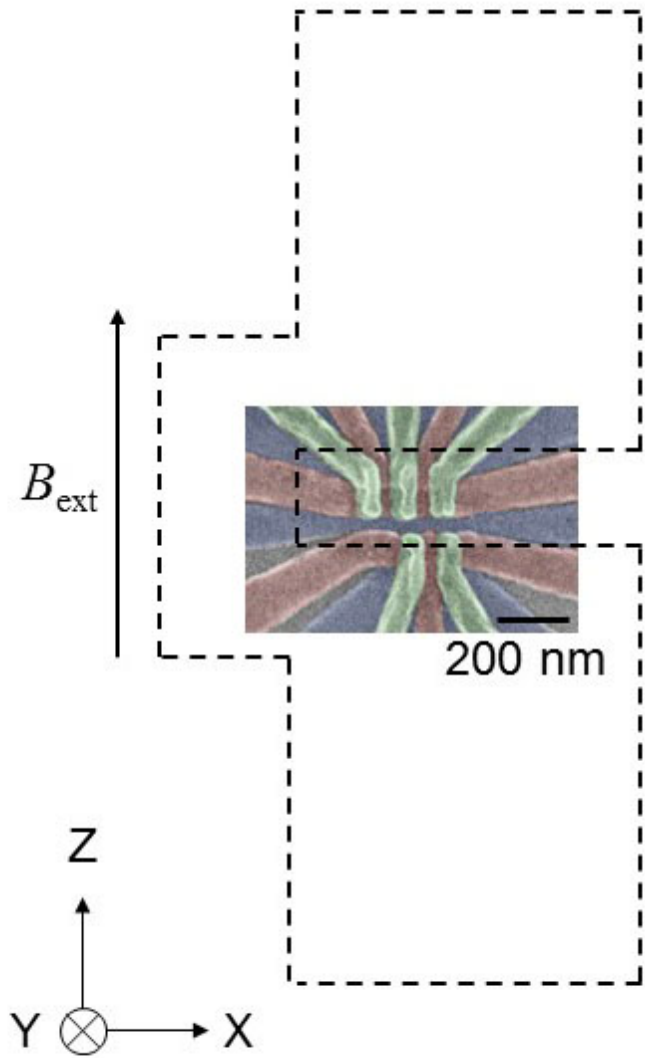

**Fig. S1**

False color scanning electron microscope image of the device with dashed lines denoting the region where Co is deposited. In these experiments an external field $B_{ext} = 0.5$ T is applied in the z-direction.

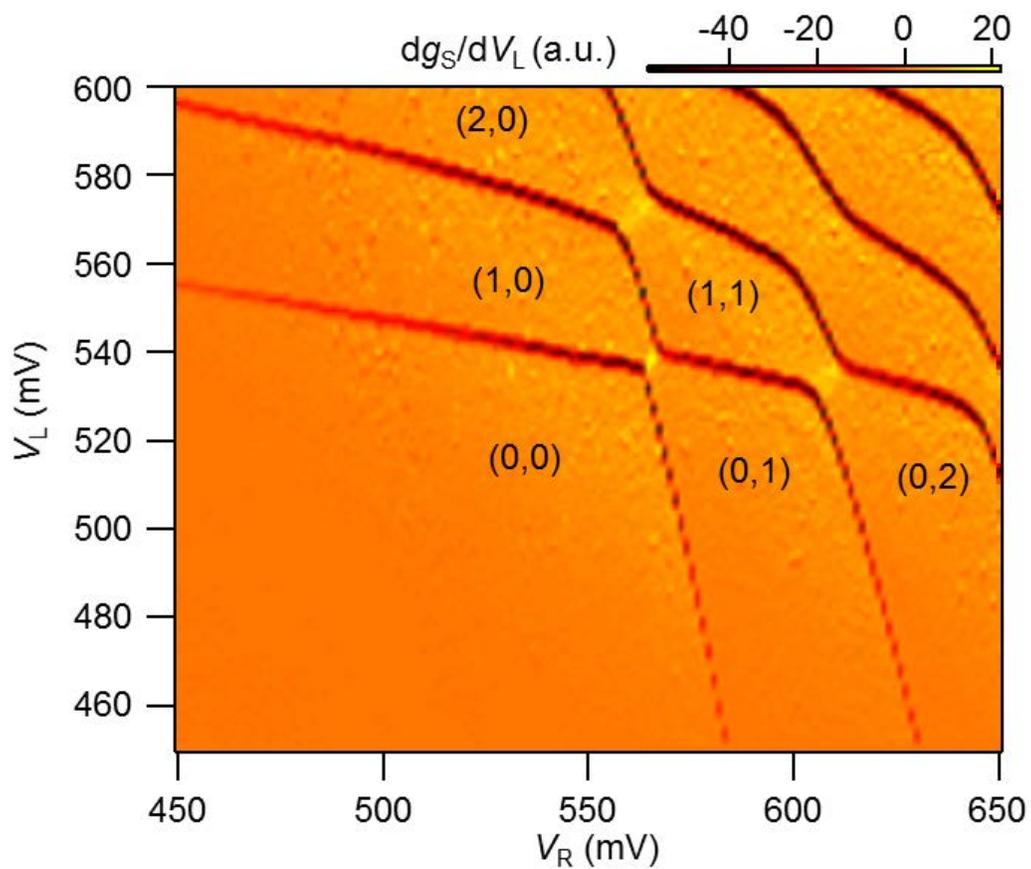

**Fig. S2**

DQD stability diagram. The derivative of $g_s$ with respect to $V_L$ is plotted as a function of $V_L$ and $V_R$. The stable charge configurations are labelled by $(N_L, N_R)$, where $N_L$ ($N_R$) is the number of electrons in the left (right) dot.

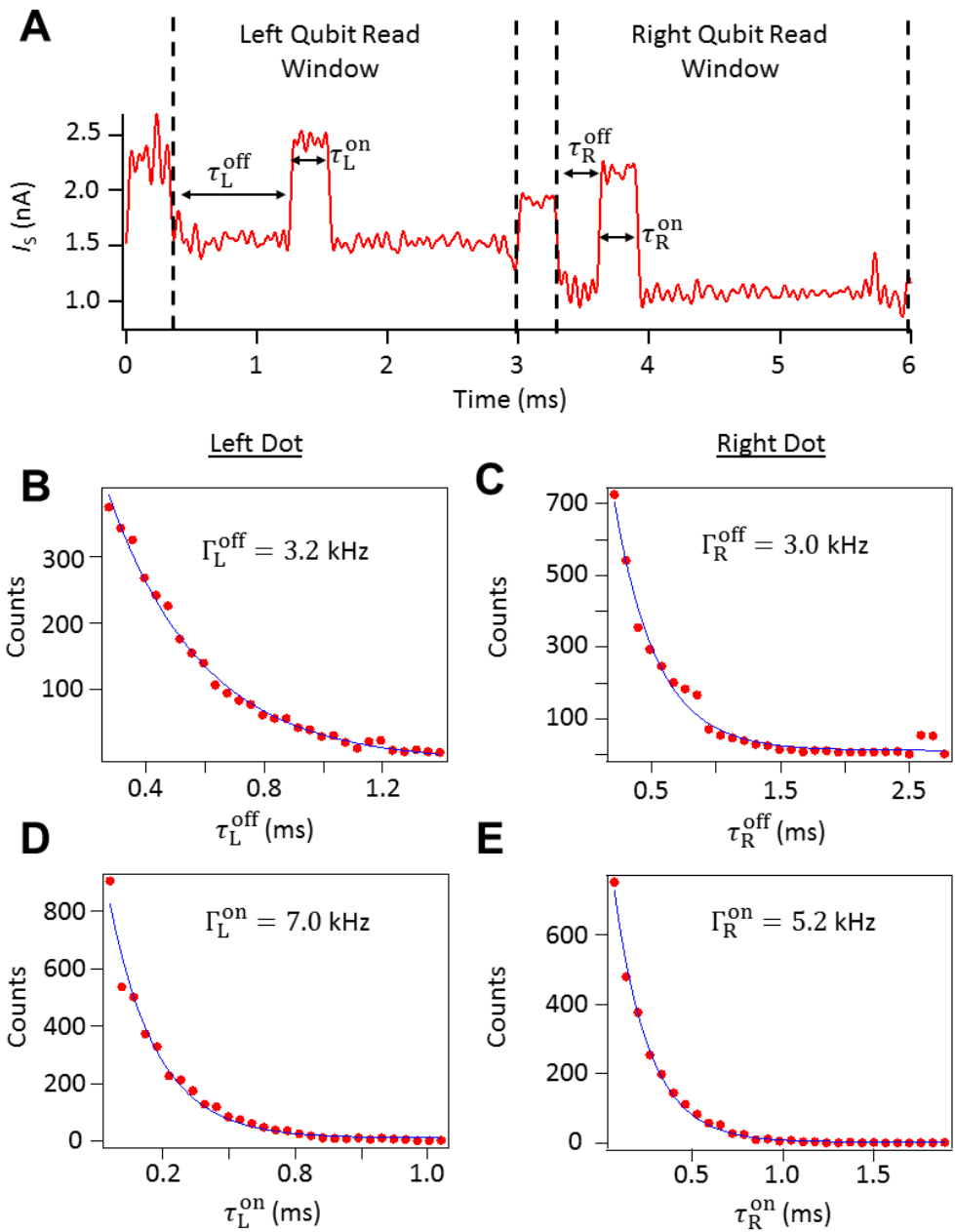

**Fig. S3**

Measuring tunneling rates during the readout process. An example single shot trace is shown in (**A**) where both electrons were detected in the spin up state. The square current blips at 1.3 and 3.7 ms are the spin signals. From these traces we can extract the time at which a spin up electron tunnels out and the time at which a spin down electron replaces it. We do this for 5,000 single shot traces and histogram the tunneling times to extract the four tunneling rates in (**B**)-(**E**).

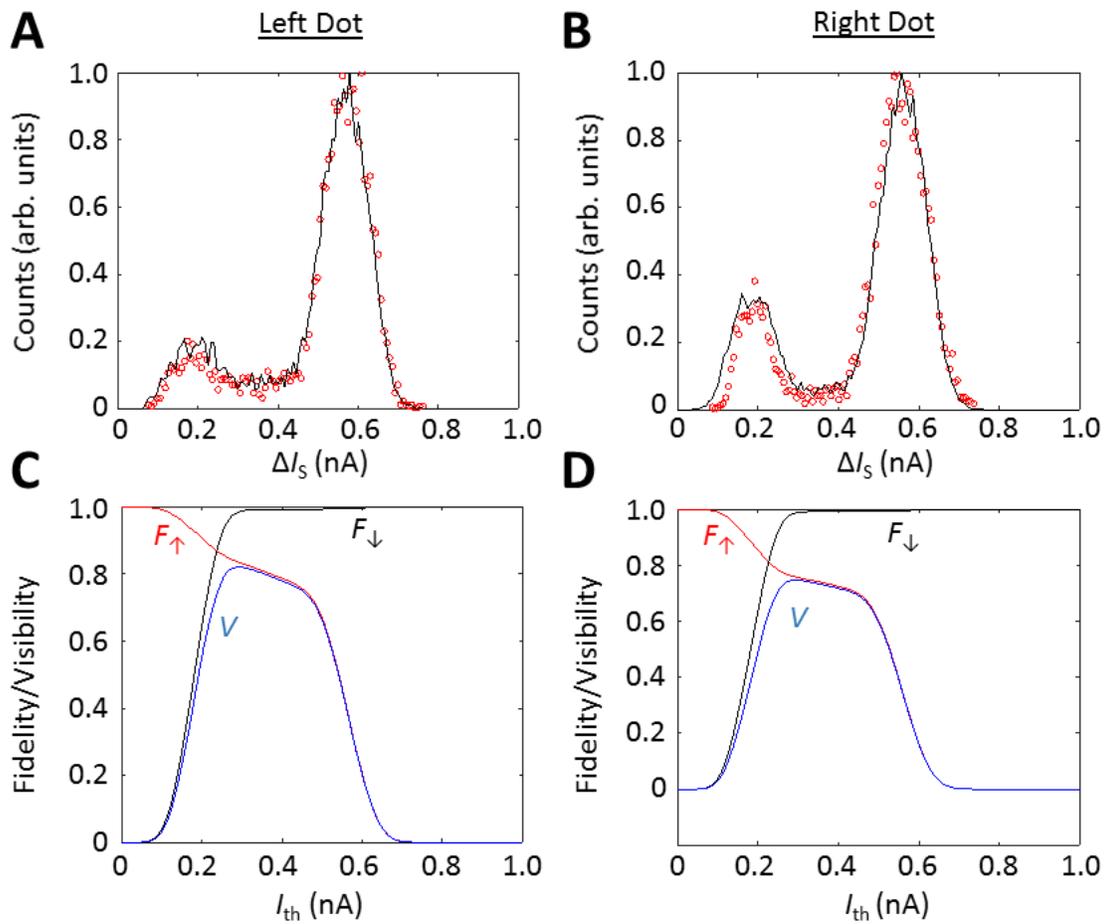

**Fig. S4**

Readout fidelities and visibilities for both quantum dots. Histograms of the peak-to-peak current $\Delta I_S$ during the read window for the left (**A**) and right quantum dots (**B**). Red dots are experimental data and black lines are simulation results. From the simulation results the fidelities for measuring the two spin states and readout visibility are plotted in (**C**) and (**D**) as a function of the current threshold $I_{th}$.

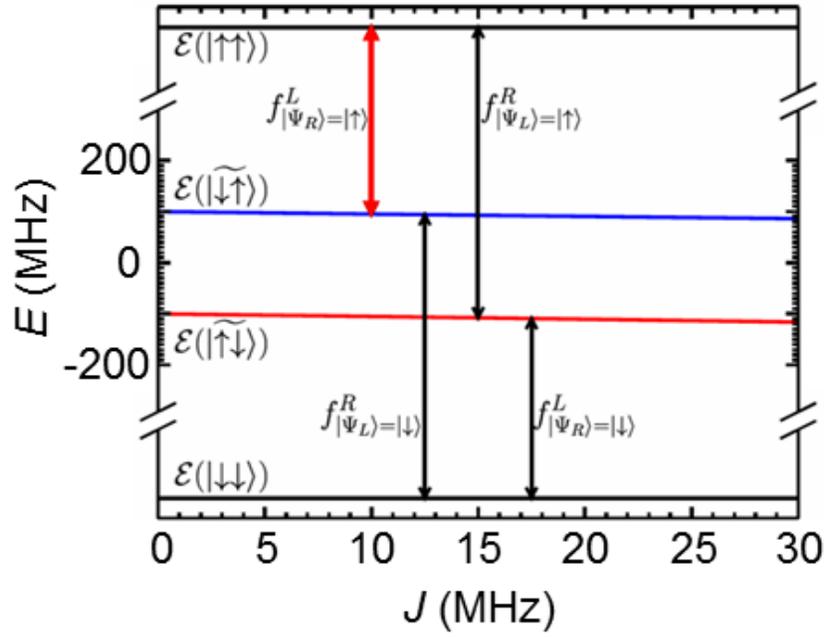

**Fig. S5**

Eigenenergies as a function of the exchange interaction $J$ for the parameters in our experiment. The eigenstates are effectively $\{|\uparrow\uparrow\rangle, |\uparrow\downarrow\rangle, |\downarrow\uparrow\rangle, |\downarrow\downarrow\rangle\}$ since the interaction between the two qubits is dominated by the large field gradient $\delta B \approx 200$ MHz $\gg J \approx 20$ MHz. The black arrows indicate the relevant transitions in our experiment.

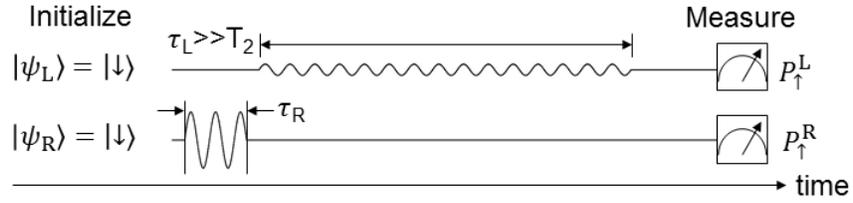

**Fig. S6**

Exchange spectroscopy pulse sequence. After initializing both qubits in spin-down a MW drive of length $\tau_R$ is used to rotate the right qubit. Then a long ($\tau_L \gg T_2$) low power MW drive is applied to the left qubit. When on resonance, the left qubit is left in a mixed state. The resonance frequency of the left qubit in Fig. 3D is seen to depend on the rotation applied to the right qubit $\tau_R$. This demonstrates that the resonance frequency of the left qubit depends on the state of the right qubit.

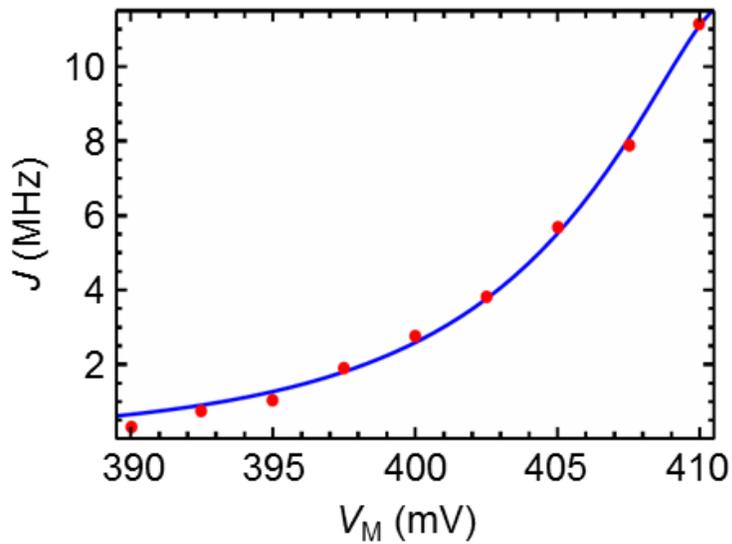

**Fig. S7**

$J$ as a function of $V_M$ (dots), measured using the pulse sequence of Fig. 3. The blue curve is a fit to the data based on tunneling through a rectangular barrier.

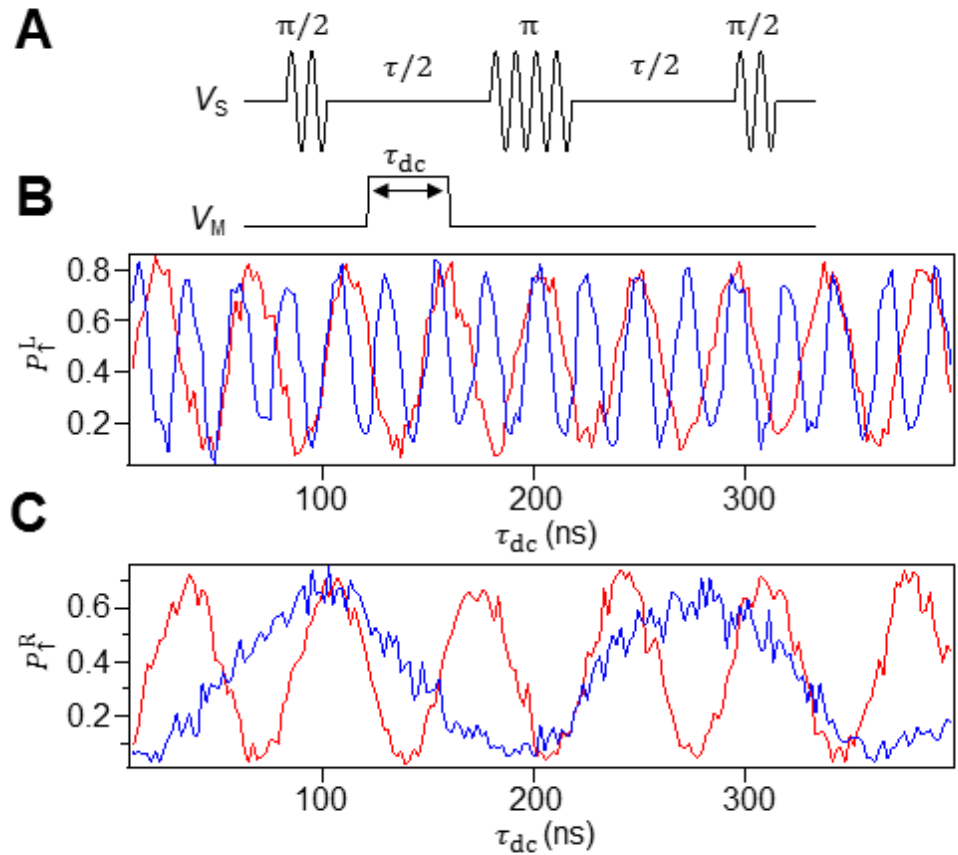

**Fig. S8**

Measuring single qubit transition frequencies with a Hahn echo sequence. (**A**) We apply a Hahn echo pulse sequence to one of the qubits with $\tau = 2$ µs. By applying a dc exchange pulse during only one period of free evolution during the echo sequence we accumulate phase proportional to the frequency shift induced by the dc exchange pulse. (**B**) $P_\uparrow^L$ as a function of $\tau_{dc}$ for the right qubit in the spin up state (blue trace) and spin down state (red trace). (**C**) $P_\uparrow^R$ as a function of $\tau_{dc}$ for the left qubit in the spin up state (blue trace) and spin down state (red trace).

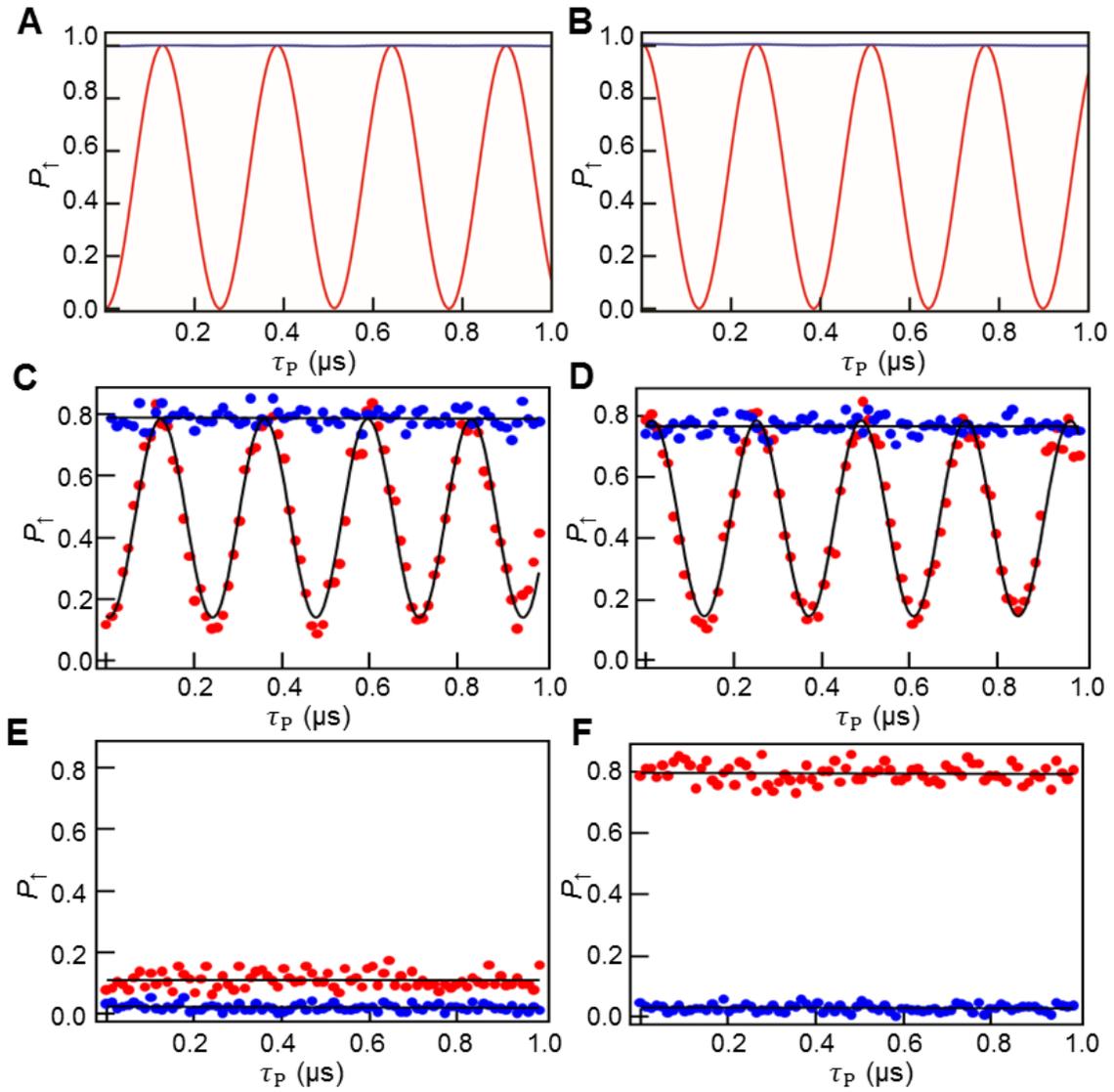

**Fig. S9**

Simulated conditional oscillations from CNOT pulse sequence. Simulated spin up probabilities for an input state $|\psi_{in}\rangle = |\downarrow\uparrow\rangle$ (**A**) and for $|\psi_{in}\rangle = |\uparrow\uparrow\rangle$ (**B**). Measured spin up probabilities for $|\psi_{in}\rangle = |\downarrow\uparrow\rangle$ (**C**), $|\psi_{in}\rangle = |\uparrow\uparrow\rangle$ (**D**), $|\psi_{in}\rangle = |\downarrow\downarrow\rangle$ (**E**), and $|\psi_{in}\rangle = |\uparrow\downarrow\rangle$ (**F**). The measured oscillation amplitudes differ from the simulations due to limited readout visibility. The black curves are least squares fits to the data.

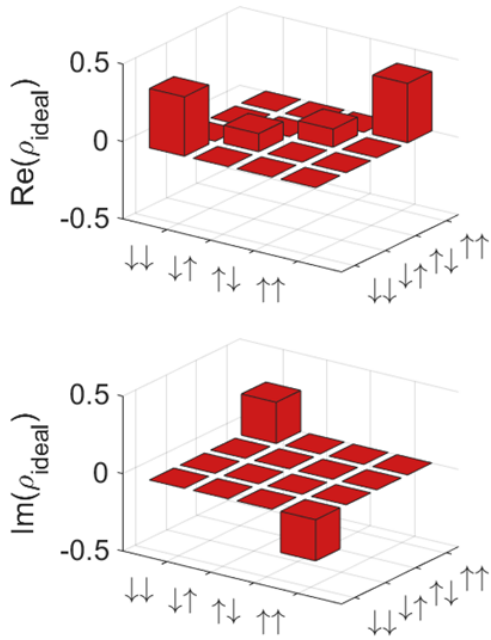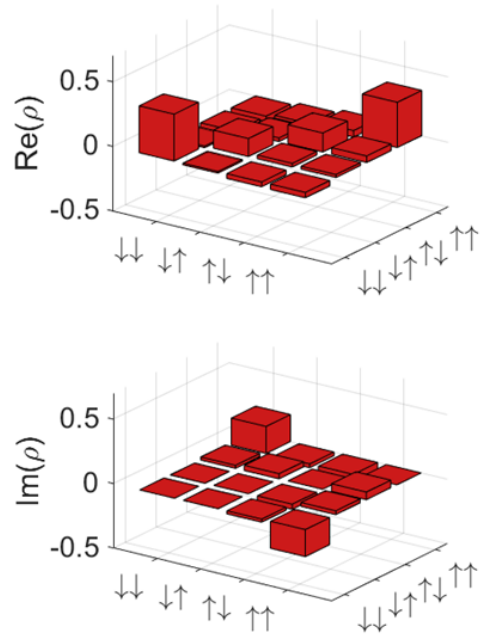

**Fig. S10**

(**A**) Expected density matrix of the ideal Bell state with the experimentally measured readout visibility. (**B**) Experimentally measured density matrix with a Bell state fidelity $F = 75\ \%$.